\documentclass[a4paper,11pt]{article}

\usepackage{graphicx}     
\usepackage{physics}      
\usepackage{mathrsfs}     
\usepackage{mathtools}    
\usepackage{tensor}       
\usepackage{subcaption}   
\usepackage{xcolor}       
 \usepackage{float}

\pdfoutput=1 

\usepackage{jcappub} 

\usepackage[T1]{fontenc} 

\title{\boldmath Particle production and Higgs reheating}

\author[a]{Aarav Shah}
\author[b]{Kanabar Jay}
\author[a]{Maxim Khlopov}
\author[c]{Oem Trivedi}
\author[d,e]{Maxim Krasnov}
\affiliation [a] {Virtual Institute of Astroparticle Physics, 75018 Paris, France}
\affiliation[b]{MG Science Institute, Gujarat University, Ahmedabad 380009, India}
\affiliation[c]{Department of Physics and Astronomy, Vanderbilt University, Nashville, TN 37235, USA}
\affiliation[d]{National Research Nuclear University MEPhI, 115409 Moscow, Russia}
\affiliation[e]{Research Institute of Physics, Southern Federal University, 344090 Rostov-on-Don, Russia}
\emailAdd{shahaarav103@zohomail.in}
\emailAdd{kanabarjaymgsi@gmail.com}
\emailAdd{khlopov@apc.in2p3.fr}
\emailAdd{oem.trivedi@vanderbilt.edu}
\emailAdd{morrowindman1@mail.ru}


\abstract{Reheating is essential for transforming the cold, vacuum dominated Universe at the end of inflation into the hot thermal bath required by the Standard Model. In many well motivated inflationary models, however, the inflaton has no direct couplings to other fields, raising the question of how the Universe becomes repopulated with particles. We address this question within the framework of geometric reheating, where energy transfer occurs purely through gravitational effects. Focusing on a Higgs inflationary scenario with a non-minimal curvature coupling $\xi \phi^2 R$, we derive the post-inflationary dynamics and compute particle production using the Bogoliubov formalism. We show that the rapid, oscillatory evolution of the curvature scalar after inflaton acts as a time dependent gravitational pump, creating scalar spectator particles even in the absence of explicit interactions. This curvature driven production mechanism provides a natural and efficient route to reheating, demonstrating that gravity alone can initiate the standard thermal history and bridge inflation with radiation domination in minimal, coupling free models of the early Universe.}

\begin{document}
\maketitle
\flushbottom

\section{Introcution}
The inflationary paradigm \cite{Guth1981InflationaryUniverse,Linde1982NewInflation,AlbrechtSteinhardt1982Cosmology,Linde1983ChaoticInflation,MukhanovChibisov1981QuantumFluctuations,Starobinsky1980NewType} provides a compelling explanation for the observed homogeneity, isotropy, and near scale invariance of the cosmic microwave background (CMB) \cite{Planck2020ResultsVI,Hu2002CMBReview,Bennett2013WMAP9}. Yet, the details on how the Universe transitions from the cold, vacuum dominated inflationary state \cite{Guth1981InflationaryUniverse,Linde1982NewInflation,AlbrechtSteinhardt1982Cosmology,Linde1983ChaoticInflation,MukhanovChibisov1981QuantumFluctuations,Starobinsky1980NewType} to the hot, radiation dominated era of the Standard Model (SM) \cite{Bezrukov2009InitialConditionsHotBB,GarciaFigueroaRubio2009,Figueroa2016SMHiggsOriginHotBB,Laine2020GWBackgroundSM} remains an open question. This transition, known as reheating \cite{KolbTurner1990,choi2024minimalproductionpromptgravitational,Dimopoulos_2018,Kofman1994Reheating,Kofman1997TowardsTheory,Greene2000FermionicPreheating,Greene1997StructureResonance,Allahverdi2010ReheatingReview,Micha2004TurbulentThermalization,Bassett2006InflationDynamicsReheating}, governs the initial conditions for all subsequent cosmic evolution and links inflationary dynamics to particle physics.
\\
\\
After inflation \cite{Guth1981InflationaryUniverse,Linde1982NewInflation,AlbrechtSteinhardt1982Cosmology,Linde1983ChaoticInflation,MukhanovChibisov1981QuantumFluctuations,Starobinsky1980NewType}, the Universe is extremely cold and dilute, with almost all of its energy stored in the coherent oscillations of the inflaton. These oscillations behave like non relativistic matter and do not automatically generate the thermal radiation bath required for Big Bang evolution. Reheating is therefore essential \cite{KolbTurner1990,choi2024minimalproductionpromptgravitational,Dimopoulos_2018,Kofman1994Reheating,Kofman1997TowardsTheory,Greene2000FermionicPreheating,Greene1997StructureResonance,Allahverdi2010ReheatingReview,Micha2004TurbulentThermalization,Bassett2006InflationDynamicsReheating}, it repopulates the Universe with particles and initiates the standard hot thermal history.
\\
\\
In the standard picture, reheating proceeds through perturbative decays or parametric resonance of the inflaton into lighter degrees of freedom \cite{Kofman1994Reheating,Kofman1997TowardsTheory,Greene2000FermionicPreheating,Greene1997StructureResonance,Allahverdi2010ReheatingReview,Micha2004TurbulentThermalization,Bassett2006InflationDynamicsReheating,Traschen:1990sw,Dolgov:1989us}. However, in many theoretically motivated models; particularly those arising from supergravity \cite{KalloshLinde202}, string theory   \cite{Mukhi2011,Wen2024,CveticEtAl2022,BerkovitsEtAl2022,Coudarchet2023,MarchesanoShiuWeigand2024}, or grand unification \cite{Georgi1974SU5,Fritzsch1975SO10,Langacker1981GUTReview}; the inflaton couples only gravitationally to other fields and, explicit couplings are either suppressed or absent \cite{OdintsovOikonomou2019,MarchesanoShiuWeigand2024,Coudarchet2023,BerkovitsEtAl2022,CveticEtAl2022,Wen2024,Mukhi2011}. In such cases, energy transfer to the SM sector must occur through gravitational or geometric effects, a mechanism collectively referred to as gravitational reheating \cite{BassettLiberati1998,HaqueMaity2022}. The study of this process is especially relevant in minimal models of the early Universe, where simplicity and predictive power are preserved by avoiding arbitrary  couplings \cite{Starobinsky1980NewType,Linde1983ChaoticInflation,BezrukovShaposhnikov2008,Kallosh2013Universality,Ema2017GravitationalReheating}.
\\
\\
An intriguing realization of this scenario is provided by Higgs inflation \cite{chowdhury2025higgsinflationparticlefactory,Rubio2019,EmaNakayama2021,Salvio2019,FumagalliMooijPostma2021}, in which the Standard Model Higgs doublet plays the role of the inflaton \cite{chowdhury2025higgsinflationparticlefactory,BezrukovShaposhnikov2008,DeSimoneHertzbergWilczek2009} through a non-minimal coupling to the Ricci scalar.  In this work, we revisit reheating such Higgs like inflationary models from a purely geometric standpoint \cite{BassettLiberati1998,HaqueMaity2022,MarkkanenRajantie2017,HashibaYokoyama2019,Giovannini2021,DomenechSasaki2021}. We consider an inflaton with a general non-minimal coupling to curvature and analyze its gravitational interaction with a set of spectator scalar fields $\chi_n$. \footnote{See \cite{ChakrabortyMaitiMaity2024} for the impact of scalar fluctuations ($\chi$) non-minimally coupled to gravity, $\xi\chi^2 R$, as a potential source of secondary gravitational waves (SGWs). }
\\
\\
In the second part of the paper, we develop the formalism of gravitational particle production using the Bogoliubov transformation of the spectator field modes \cite{Ford1987,BirrellDavies1982,Fulling1989,ChakrabortyCleryHaqueMaityMambrini2025,ChakrabortyMaityMondal2025}. By solving the mode equation in a Friedmann Lema\^{i}re Robertson Walker (FLRW) background \cite{WeinbergCosmology2008,Baumann2022CosmologyBook,Mukhanov2005PhysicalFoundations,Dodelson2003ModernCosmology}, we determine the occupation numbers, comoving number densities, and the out of equilibrium corrections induced by finite decay rates \cite{HaqueMaity2022}. This framework enables a quantitative connection between the curvature induced creation of particles and the subsequent thermalization of the universe.
\\
\\
The remainder of this paper is organized as follows. In Sec.\ref{Section 2}, we present the theoretical setup for Higgs reheating, introducing the action, field equations, and the relevant energy exchange relations. In Sec.\ref{Section 3}, we analyze gravitational particle production and its contribution to reheating. Finally, Sec.\ref{Conclusions} summarizes our conclusions and discusses the implications of geometric reheating for early universe cosmology.

\section{Higgs Reheating} \label{Section 2}
\begin{figure}[H]
    \centering
    
    \begin{subfigure}{0.45\textwidth}
        \includegraphics[width=\linewidth]{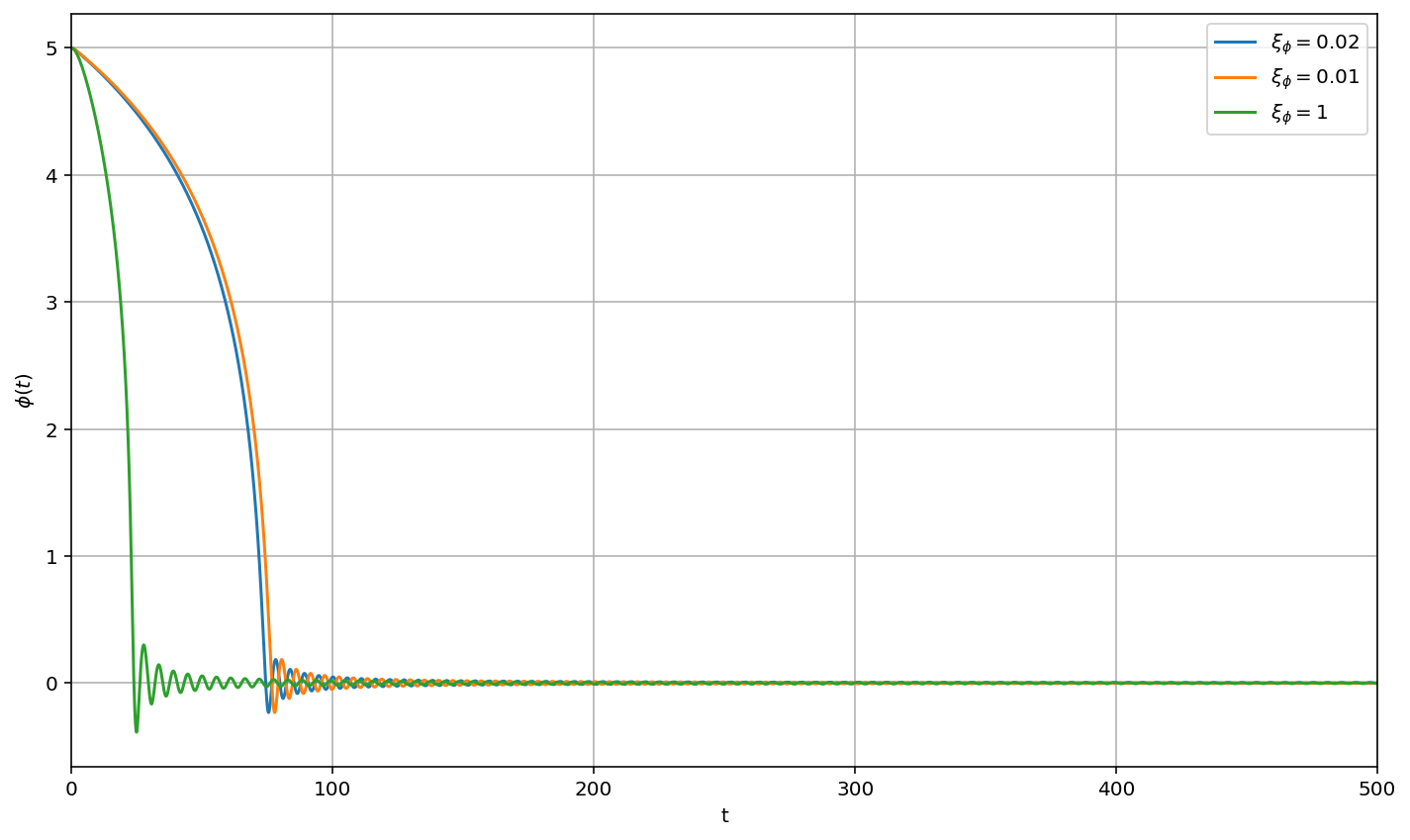}
        \caption{Evolution of ${\phi}$ in a $\alpha$ attractor model for different values of $\xi$.}
        \label{fig:1a}
    \end{subfigure}
    \hfill
    \begin{subfigure}{0.45\textwidth}
        \includegraphics[width=\linewidth]{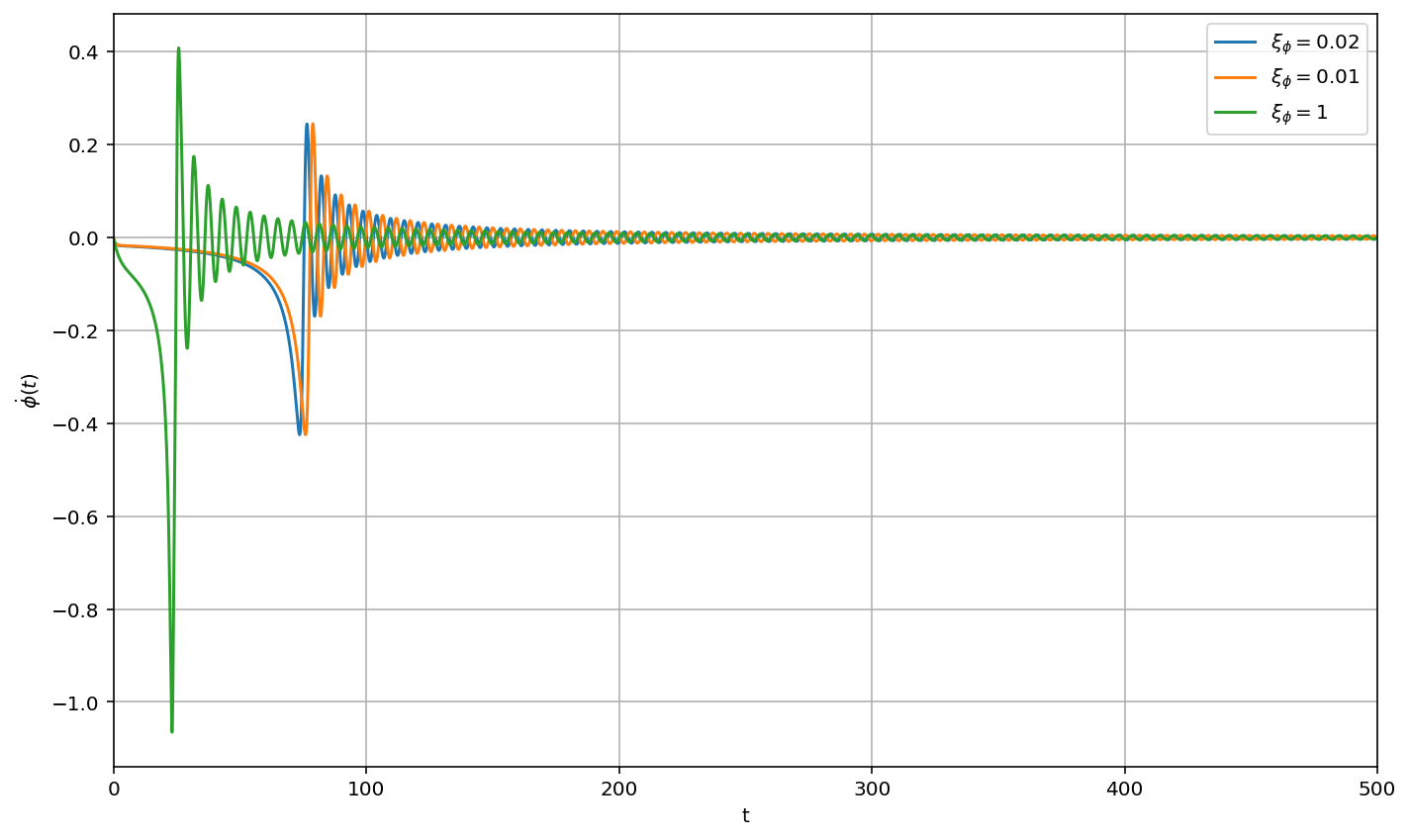}
        \caption{Evolution of $\dot{\phi}$ in a $\alpha$ attractor model for different values of $\xi$.}
        \label{fig:1b}
    \end{subfigure}

    \caption{$\xi=0.02$ (blue), $\xi=0.01$ (orange) and $\xi=1$ (green).}
    \label{fig:1}
\end{figure}

Reheating \cite{KolbTurner1990,choi2024minimalproductionpromptgravitational,Dimopoulos_2018,Kofman1994Reheating,Kofman1997TowardsTheory,Greene2000FermionicPreheating,Greene1997StructureResonance,Allahverdi2010ReheatingReview,Micha2004TurbulentThermalization,Bassett2006InflationDynamicsReheating,BassettLiberati1998,HaqueMaity2022} marks the crucial bridge between the end of inflation and the beginning of the hot Big Bang. After inflation, the universe is cold and dominated by the coherent oscillations of the inflaton . During reheating, this stored vacuum energy must be converted into ordinary particles; photons, baryons, scalar spectator particles $\psi$ (produced via gravitional particle production as highlighted in Sec.\ref{Section 3}) and possibly dark matter to establish the thermal bath that later evolves into the Standard Model plasma \cite{DeSimoneHertzbergWilczek2009}. The precise nature of this energy transfer determines the starting point of standard cosmology and links inflationary dynamics to high energy particle physics \cite{HaqueMaity2022}.
\\
\\
In conventional scenarios, reheating occurs because the inflaton is explicitly coupled to lighter fields through interaction terms such as $g\phi^2\chi^2$ or Yukawa type couplings \cite{BassettLiberati1998}. As the inflaton oscillates around the minimum of its potential, these couplings allow perturbative decays or resonant preheating, efficiently producing quanta of other fields \cite{Kofman1994Reheating,Kofman1997TowardsTheory,Greene2000FermionicPreheating,Greene1997StructureResonance,Allahverdi2010ReheatingReview,Micha2004TurbulentThermalization,Bassett2006InflationDynamicsReheating,Traschen:1990sw,Dolgov:1989us}. However, such mechanisms depend on model specific choices of coupling constants whose magnitudes are often ad hoc. In many ultraviolet motivated models such as those emerging from supergravity \cite{KalloshLinde202}, string theory \cite{Mukhi2011,Wen2024,CveticEtAl2022,BerkovitsEtAl2022,Coudarchet2023,MarchesanoShiuWeigand2024}, or grand unified constructions \cite{Georgi1974SU5,Fritzsch1975SO10,Langacker1981GUTReview}, the inflaton may interact only gravitationally with the visible sector. In this case, particle creation must proceed through the geometry of spacetime itself. We refer to this minimal, coupling free scenario as geometric reheating \cite{BassettLiberati1998,HaqueMaity2022}.
\\
\\
In reheating models without explicit couplings to other fields \cite{KolbTurner1990,choi2024minimalproductionpromptgravitational,Dimopoulos_2018,Kofman1994Reheating,Kofman1997TowardsTheory,Greene2000FermionicPreheating,Greene1997StructureResonance,Allahverdi2010ReheatingReview,Micha2004TurbulentThermalization,Bassett2006InflationDynamicsReheating}, the only dynamical quantity capable of transferring energy is the spacetime curvature. During the oscillatory phase of the inflaton, the Ricci scalar $R(t)$ also oscillates rapidly. These oscillations modulate the effective mass of any non-minimally coupled field and act as a pump of energy from the inflaton sector into other degrees of freedom. This is the essence of geometric reheating, gravity itself performs the energy transfer \cite{BassettLiberati1998,HaqueMaity2022}.
\\
\\
To illustrate this idea, one may treat the Standard Model Higgs as a prototype scalar field that drives or participates in inflation \cite{chowdhury2025higgsinflationparticlefactory,BezrukovShaposhnikov2008}. For cosmological purposes, the Higgs doublet can be represented by a single real scalar field $\phi$ with an action of the form

\begin{equation}
\begin{split}
        S=\int d^4x\sqrt{-g}\left[-\frac{1}{2}(\partial\phi)^2-V(\phi)\right.\\\left.-\frac{1}{2}\sum_\nu^Nm^2_\nu\chi_\nu^2 -\frac{1}{2}\xi\phi^2R-\frac{1}{2}\sum_\nu^N\xi_\nu\chi_\nu^2R+\frac{1}{2}\sum_\nu^N(\partial\chi_\nu)^2\right].
\end{split}
\end{equation}
Here $\phi$ plays the role of the inflaton which couples to curvature through the parameter $\xi$, while the fields $\chi_\nu$ act as spectators, vital for reheating \cite{BassettLiberati1998}, that couple to curvature through their non-minimal parameters $\xi_\nu$ . Because the only communication between the sectors is via the spacetime curvature $R$, any particle production arises purely from gravitational effects.
\\
\\
The non-minimal coupling $\xi \phi^2 R$ effectively modifies how the inflaton interacts with the spacetime curvature. When $\phi (t)$ oscillates after inflation, this coupling alters the effective gravitational response of the system. Because the curvature scalar $R(t)$ depends on both the Hubble parameter and its time derivative, the oscillations of $\phi(t)$ induce corresponding oscillations in $R(t)$. These rapid variations in curvature act as a time dependent gravitational driving force, which can transfer energy into any field that is sensitive to curvature,  such as non-minimally coupled spectator fields.
\\
\\
If $\phi$ is identified with the Higgs field, its potential is
\begin{equation}
    V(\phi)=\frac{\lambda}{4}(\phi^2-\nu^2)^2
\end{equation}
where $\nu=246 \rm{GeV}$ and $\lambda =\frac{m_h^2}{2\nu^2}\simeq 0.1$. Matching the inflationary amplitude observed in the CMB typically requires a large non-minimal coupling $\xi\sim5000\sqrt{\lambda}$. 
Such a value suppresses the effective Planck scale to $\frac{M_{Pl}}{\xi}\sim10^{14}\rm{GeV}$, close to the inflationary Hubble rate $H_{inf}\sim 10^{13}\rm{GeV}$. This near coincidence suggests that perturbative unitarity could break down during Higgs inflation.
\\
\\
However, this conclusion changes once the renormalization group (RG) running of $\lambda$ is taken into account. At high energies, the Higgs self coupling decreases and may even become slightly negative, while new heavy states can modify its trajectory altogether. Because $\xi$ and $\lambda$ are connected through the CMB normalization, reducing $\lambda$ allows proportionally smaller values of $\xi$. Realistic models can therefore operate safely with $\xi \sim 10$, avoiding the unitarity issue while still reproducing the inflationary spectrum. In what follows, $\xi$ will be treated as a free but phenomenologically constrained parameter.\footnote{See \cite{chowdhury2025higgsinflationparticlefactory} for more information}
\\
\\
Throughout this work, the inflaton $\phi$ is treated as a generic real scalar with Higgs type non-minimal coupling to curvature. The identification with the Standard Model Higgs is not assumed unless explicitly stated.
\\
\\
Having established the theoretical setup and clarified how the inflaton-curvature coupling shapes the overall framework, we now turn to the dynamical evolution of the system. To understand how energy stored in the inflaton is transferred to radiation and spectator fields, it is essential to describe the background cosmology and derive the governing equations of motion. 
\\
\\
In what follows, we shall assuming a Friedmann Lema\^{i}tre Robertson Walker background \cite{Baumann2022CosmologyBook,WeinbergCosmology2008,Mukhanov2005PhysicalFoundations,Dodelson2003ModernCosmology} 
\begin{equation}
    ds^2=-dt^2+a^2(t)(dx^2+dy^2+dz^2),
\end{equation}
and use the convention
\begin{equation}
    R=-6(\dot{H}+2H^2).
\end{equation}
We adopt $\xi_\nu$ are equal to $-\xi$ (we shall restrict our considerations to positive values of $\xi$ in our numerical evaluations, so as to keep all $\xi_\nu$'s negative\footnote{In the convention $R=+6(\dot{H}+2H^2)$, this translates to having $\xi$ negative and having all $\xi_\nu$'s positive. To better understand the phenomenological consequences of assuming $\xi_\nu$ to be positive, see the thesis of Brain R. Greene \cite{Greene_1997}.})  as a benchmark case corresponding to symmetric but opposite sign curvature couplings between the inflaton and matter sectors. In more general EFT settings, $\xi_{\nu}$ may differ from $\xi$, a case we leave for future work.
\\
\\
In this background, the inflation obeys
\begin{equation}
\label{inflation field}   
\ddot{\phi}+3H\dot{\phi}-6\xi\phi(H^2+2\dot{H})+\frac{dV(\phi)}{d\phi}+\Gamma^T_\phi \rho_\phi (1+\omega_\phi)=0.
\end{equation}
where $H=\frac{\dot{a}}{a}$ is the Hubble parameter, $\omega_{\phi}$ is the inflationary equation of state parameter and $\Gamma^T_{\phi}$ denotes an effective decay or dissipation rate that captures the transfer of energy from the inflaton to the background radiation and dark matter.
\\
\\
The modified Friedmann equation takes the form
\begin{equation}
\begin{split}
       3M_{Pl}^2H^2+\xi\phi^2\left(H^2+6H\frac{d(log\phi)}{dt}\right)=\frac{\dot{\phi}^2}{2}+V(\phi)\\+\frac{1}{2}\sum_\nu^Nm^2_\nu\chi_\nu^2-\frac{\xi}{2}\sum_\nu^N\chi_\nu^2R.
\end{split}
\label{modified Friedmann equation}
\end{equation}
The extra term coupled to $\xi$ effectively alters the gravitational coupling and hence modifies the expansion history compared with minimally coupled inflaton. To better understand the mechanism, along with the model independent consideration, we also consider $\alpha$ attractor model \cite{KalloshLinde202,OdintsovOikonomou2019,ShahalamMyrzakulov2020} with inflationary potential
\begin{equation}\label{inflation potential}
    V(\phi)=\Lambda^4\left[1-e^{-\sqrt{\frac{2}{3\alpha}}\phi/M_{Pl}}\right]^{2n}.
\end{equation}
Using eqns.(\ref{inflation field},\ref{modified Friedmann equation},\ref{inflation potential}), we plot the behavior of inflaton $\phi$ and its velocity $\dot{\phi}$ in the $\alpha$ attractor model (with $n=1,\alpha=1$) for various values of the non-minimal coupling $\xi$ in the Fig.\ref{fig:1}. 

From fig.\ref{fig:1a}, we observe that the inflaton amplitude decreases monotonically with time, indicating the onset of reheating as the field’s coherent oscillations redshift away. Larger values of $\xi$ lead to a more rapid decay of $\phi$, consistent with stronger coupling to curvature and thus a faster transfer of energy from the inflaton to the background geometry. From fig.\ref{fig:1b}, we can see the evolution of $\dot{\phi}$ revealing oscillations about zero that gradually diminish in amplitude. These oscillations correspond to the inflaton’s periodic motion around the minimum of its potential, with their damping rate governed by the Hubble expansion and the effective friction induced by the non-minimal coupling. A larger $\xi$ results in higher frequency oscillations at early times and a quicker attenuation, further supporting the interpretation that geometric effects enhance the efficiency of reheating. 
\\
\\
Each spectator $\chi_k$ satisfies
\begin{equation}
\label{spectaor chi eqn}
    \ddot{\chi}^\nu_k+3H\dot{\chi}^\nu_k+(m_\nu^2-\xi R)\chi^\nu_k=0,
\end{equation}
whose behavior encodes the curvature induced particle creation.
\\
\\
Writing the equation in canonical form gives    
\begin{equation}
    \frac{d(a^{3/2}\chi_k)}{dt}+(m_\nu^2+3H^2+2\dot{H})\chi_k=0,
\end{equation}
which resembles a harmonic oscillator with a time dependent frequency controlled by the background expansion. To better understand the relationship between  $\chi_k$ and $\xi$, we plot the evolution of $\chi_k$  against different values of  $\xi$ in fig.\ref{fig:2}.

\begin{figure}[H]
    \centering

    \begin{subfigure}[b]{0.48\textwidth}
        \includegraphics[width=\textwidth]{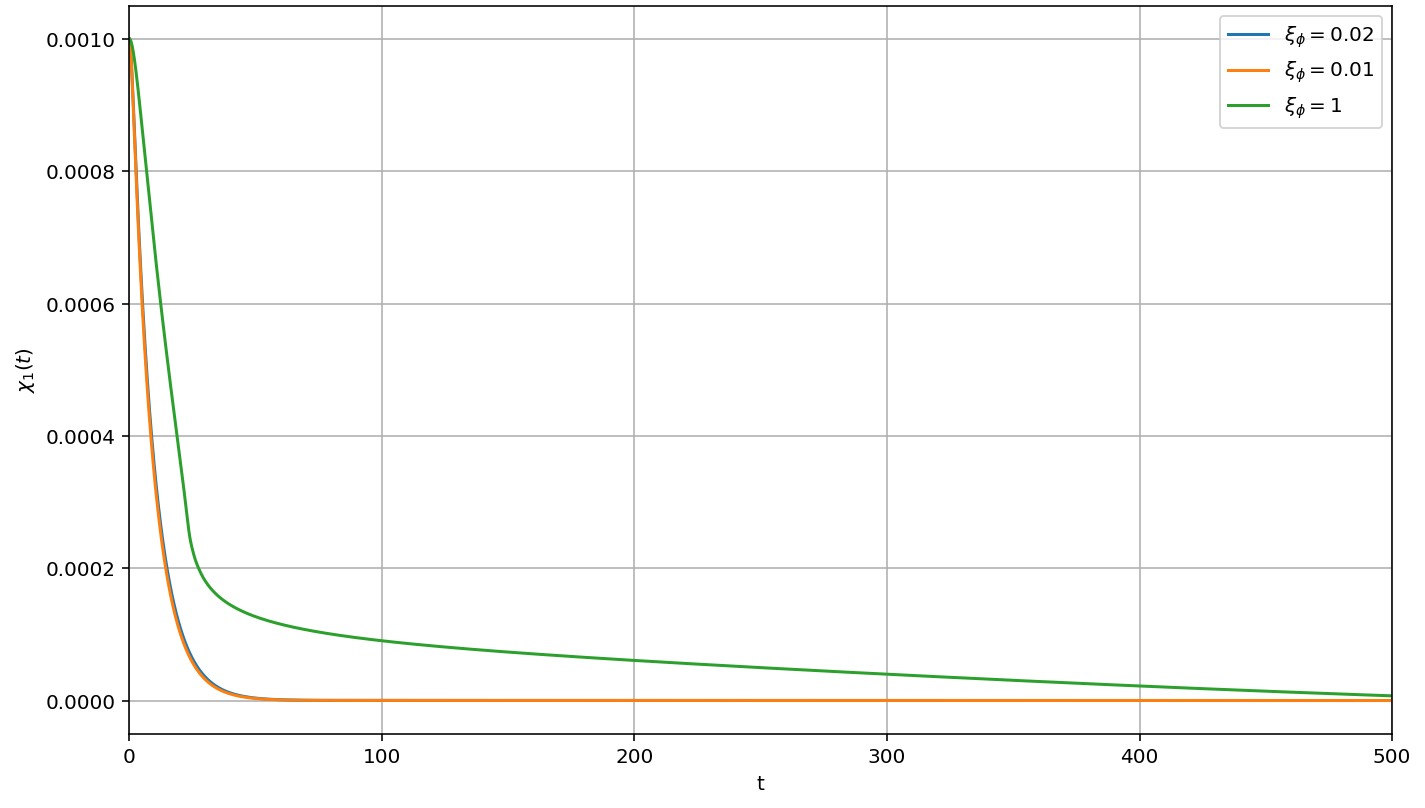}
        \caption{Evolution $\chi_1$ (set at initial value $\chi_1=10^{-3}$) for different values of $\xi$.}
        \label{fig:2_a}
    \end{subfigure}
    \hfill
    \begin{subfigure}[b]{0.48\textwidth}
        \includegraphics[width=\textwidth]{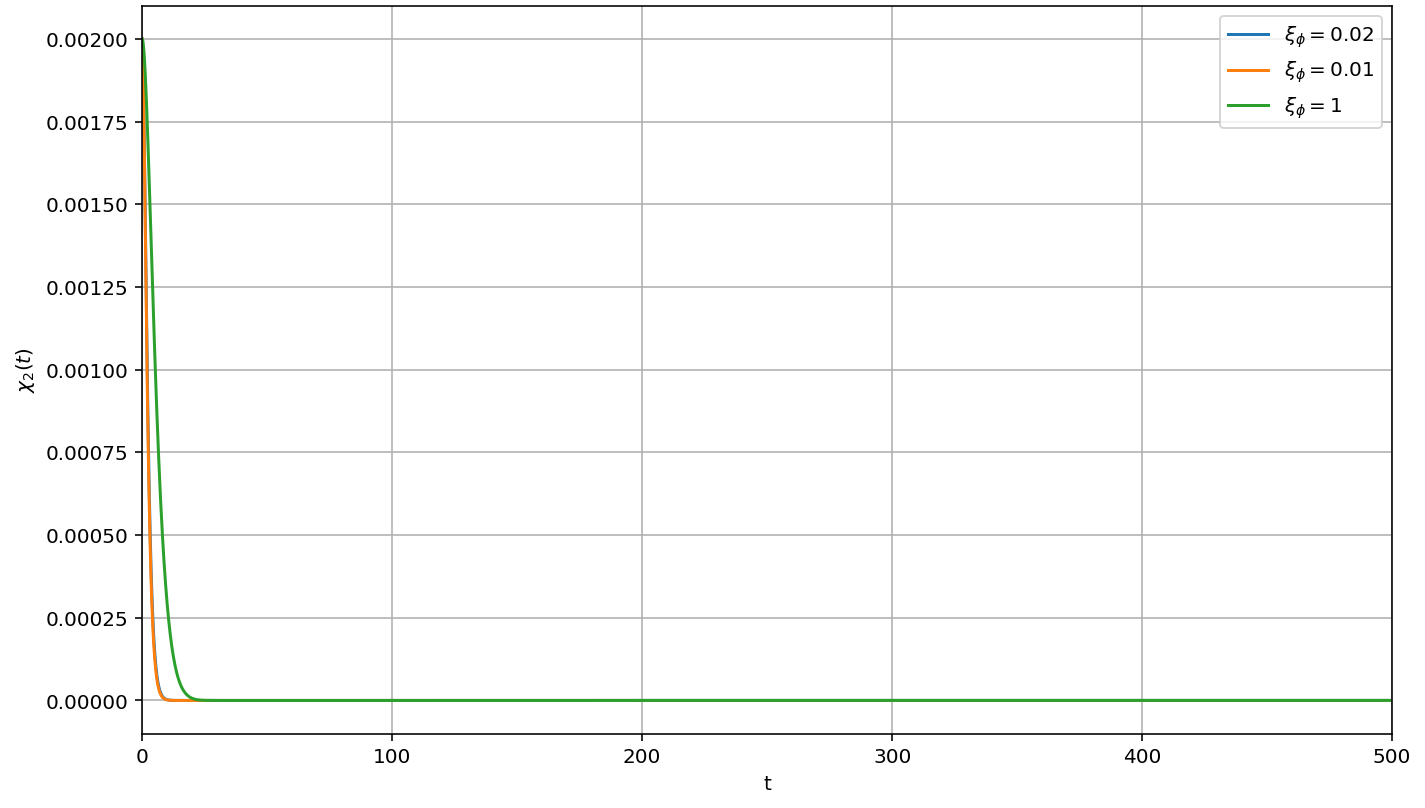}
        \caption{Evolution $\chi_2$ (set at initial value $\chi_1=2\times 10^{-3}$) for different values of $\xi$.}
        \label{fig:2_b}
    \end{subfigure}

    \caption{$\xi=0.02$ (blue), $\xi=0.01$ (orange) and $\xi=1$ (green).}
    \label{fig:2}
\end{figure}
\begin{figure}[H]\ContinuedFloat
    \centering

    \begin{subfigure}[b]{0.6\textwidth}
        \includegraphics[width=\textwidth]{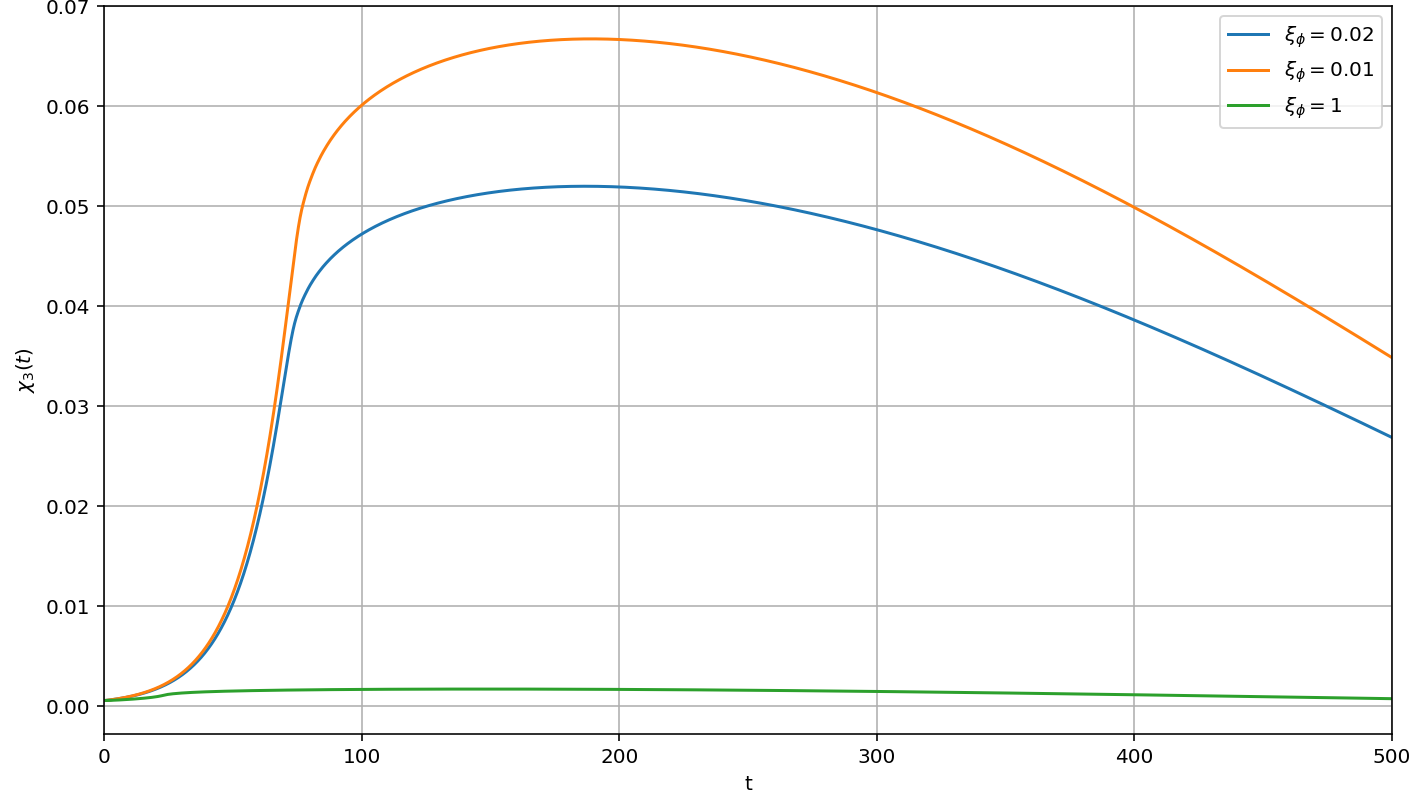}
        \caption{Evolution $\chi_3$ (set at initial value $\chi_1=5\times 10^{-5}$) for different values of $\xi$.}
        \label{fig:2_c}
    \end{subfigure}

\end{figure}

The evolution of the spectator fields shown in Figs.\ref{fig:2_a}, \ref{fig:2_b} and \ref{fig:2_c} demonstrates that different values of the spectator field $\chi_k$ can exhibit qualitatively distinct dynamical behavior, even when they share the same mass and curvature coupling. This difference arises from their initial values which determine how each field responds to the subsequent curvature driven background evolution. Although all spectator fields satisfy the same equation of motion (eqn.\eqref{spectaor chi eqn}), their trajectories in phase space are set by their initial amplitudes and velocities.
\\
\\
After inflation, the Ricci scalar undergoes rapid oscillations sourced by the inflationary dynamics. Depending on its initial configuration, a spectator field may encounter these curvature oscillations while being effectively overdamped, leading to a smooth and monotonic decay, or while behaving as an under-damped oscillator, resulting in coherent oscillations whose amplitudes gradually decrease due to Hubble friction. Consequently, even small differences in the initial values can be amplified by the time dependent curvature background, producing markedly different late time evolutions among otherwise identical spectator fields.
\\
\\
In addition to the sensitivity in initial conditions, the non minimal coupling  $\xi$ plays a crucial role in controlling the overall amplification of the spectator fields. For larger values of $\xi$, the curvature induced mass term dominates the dynamics, increasing the effective restoring force and enhancing the damping of $\xi_i$. As a result, larger $\xi$ leads to a systematic suppression of the spectator field amplitude in all cases, irrespective of whether the field undergoes oscillatory or overdamped evolution. This behavior reflects the increasingly efficient transfer of energy from the spectator sector into the background geometry as the curvature coupling is strengthened.
A larger $\xi$ strengthens the coupling between the inflaton and curvature, which effectively increases the friction term in the inflationary equation of motion. Physically, this friction represents energy draining from the inflaton to the gravitational sector. Thus, for larger $\xi$, the inflaton loses energy more rapidly, its oscillations damp faster, and reheating proceeds more efficiently.
\\
\\
Having outlined the background evolution and field equations, we now turn to the mechanisms that govern the transfer of energy from the inflaton to radiation. This allows us to quantify how efficiently the non-minimal coupling can reheat the universe once inflation ends.
\\
\\
The radiation energy density $\rho_r$ evolves according to
\begin{equation}
     d(\rho_rA^4)=\Gamma_{\phi\phi\rightarrow RR}^{Rad}\rho_{\phi}(1+\omega_\phi)\frac{A^3dA}{H}.
\end{equation}
where $A=\frac{a}{a_{end}}$ is the normalized scale factor and $\Gamma^{\rm{Rad}}_{\phi\phi\rightarrow RR}$ characterizes the decay rate of the inflaton to the background radiation \footnote{In particular, $\Gamma^T_{\phi}=\Gamma^{\rm{Rad}}_{\phi\phi\rightarrow RR}+\Gamma^{\rm{DM}}_{\phi\phi\rightarrow YY}$ where, $\Gamma^{\rm{DM}}_{\phi\phi\rightarrow YY}$ is the rate of decay from the inflaton to dark matter. }. The Suffix “end” corresponds to the end of inflation.  This expression shows that radiation production grows when the inflaton energy density is high and when the background expands slowly.
\\
\\
The inflaton energy density itself can be approximated by
\begin{equation}
\begin{split}
\rho_{\phi}=A^{-3(1+\omega_\phi)}\left[3M_{Pl}^2H_{end}^2
+\xi\phi_{end}^2\left(H_{end}^2+\frac{d(log\phi)}{dt}\Bigg|_{end}\right)\right.\\\left.-\frac{1}{2}\sum_{\nu}^Nm_\nu^2\chi^2_{\nu,end}+\frac{\xi}{2}\sum_{\nu}^N \chi_{\nu,end}^2R_{end}\right] .
\end{split}
\end{equation}
Substituting this into the previous relation yields the total radiation growth rate,

\begin{equation}
       d(\rho_rA^4)=\frac{\Gamma_{\phi\phi\rightarrow RR}^{Rad}A^{-3(1+\omega_\phi)}}{H}\left[\begin{aligned}
       3M_{Pl}^2H_{end}^2.+\xi\phi_{end}^2\left(H_{end}^2+\frac{d(log\phi)}{dt}\Bigg|_{end}\right)-\frac{1}{2}\sum_{\nu}^Nm_\nu^2\chi^2_{\nu,end}\\+\frac{\xi}{2}\sum_{\nu}^N \chi_{\nu,end}^2R_{end}
       \end{aligned}\right]. 
\end{equation}
Expressing the result in terms of the temperature through $\beta=\frac{\pi^2g_{*}}{30}$ and the recombination temperature $T_{rec}$ gives
\begin{equation}
4\beta T^3A^3d(T_{reheating}A)=\frac{\Gamma_{\phi\phi\rightarrow RR}^{Rad}A^{-3(1+\omega_\phi)}}{H}\left[\begin{aligned}3M_{Pl}^2H_{end}^2.+\xi\phi_{end}^2\left(H_{end}^2+\frac{d(log\phi)}{dt}\Bigg|_{end}\right)\\-\frac{1}{2}\sum_{\nu}^Nm_\nu^2\chi^2_{\nu,end}+\frac{\xi}{2}\sum_{\nu}^N \chi_{\nu,end}^2R_{end}
\end{aligned}
\right].
\end{equation}

These relations capture the essence of Higgs driven geometric reheating. The production of radiation depends sensitively on the non-minimal coupling $\xi$, the effective equation of state parameter $\omega_{\phi}$, and the time variation of the Hubble rate $H(t)$. A larger $\xi$ enhances the curvature coupling and can accelerate reheating, while the expansion rate governs how quickly the energy density in the inflaton redshifts away. In the next section, we examine the complementary quantum field theoretic picture in which particle creation arises from the non adiabatic evolution of field modes in curved spacetime, a process known as gravitational particle production.

\section{Gravitational Particle Production} \label{Section 3}
When inflation ends, the rapid change in the cosmic expansion rate makes spacetime itself a source of particle creation \cite{Alsarraj_2021}. Even in the absence of direct couplings between the inflaton and other fields, the time dependent background curvature can excite quantum fluctuations and populate real particles \cite{chowdhury2025higgsinflationparticlefactory}. This process, known as gravitational particle production, is a generic consequence of quantum field theory in curved spacetime \cite{BirrellDavies1982,Fulling1989}. 
\\
\\
To describe this mechanism, consider a light spectator scalar field $\psi$ evolving in a Friedmann Lema\^{i}tre Robertson Walker (FLRW) background \cite{Baumann2022CosmologyBook,Mukhanov2005PhysicalFoundations,Dodelson2003ModernCosmology,WeinbergCosmology2008}. The dynamics of this field are governed by the action
\begin{equation}
\begin{split}
    S[\psi(\eta,\textbf{x})]=\int d\eta d^3\textbf{x} \left[\frac{1}{2}a^2(\partial_\eta \psi)^2-\frac{1}{2}a^2(\nabla\psi)^2\right.\\\left.-\frac{1}{2}a^4m^2\psi^2+\frac{1}{2}a^4\bar{\xi}R\psi^2\right]
\end{split}
\end{equation}
where $\eta$ is conformal time, $a(\eta)$ is the scale factor,$m$ is the mass of the spectator, and $\bar{\xi}$ denotes its non-minimal coupling to curvature.
\\
\\
From this action, one can identify the effective mass to be 
\begin{equation}
    m_{eff}^2=m^2+\frac{1}{6}(1-6\bar{\xi})a^2(\eta)R(\eta).
\end{equation}
When $R(t)$ varies rapidly, the effective mass of the spectator field also varies rapidly. If this variation occurs faster than the field can adjust (violating adiabaticity), the field is forced to excite quanta. This is the gravitational equivalent of particle production in a time dependent medium, the curvature oscillations pump energy into $\chi$ particles.
\\
\\
To study this effect quantitatively, we begin by quantizing a test scalar field in an expanding Friedmann Lema\^{i}tre Robertson Walker background and decomposing it into Fourier modes \cite{BirrellDavies1982,Fulling1989}. This allows us to track how the curvature of spacetime modifies the evolution of individual momentum modes. The quantum field $\psi$ can be decomposed into Fourier modes as
\begin{equation}
    \hat{\psi}(\eta,\textbf{x})=\int \frac{d^3\textbf{k}}{(2\pi)^3}\left[\hat{a}_{\textbf{k}}\psi_{\textbf{k}}(\eta)e^{i\textbf{k}\cdot \textbf{x}}+\hat{a}_{\textbf{k}}^{\dagger}\psi_{\textbf{k}}^*e^{-i\textbf{k}\cdot \textbf{x}}\right],
\end{equation}
where $a_{\textbf{k}}$ and $a^{\dagger}_{\textbf{k}}$ are annihilation and creation operators, respectively.
\\
\\
Each mode function $\psi_{\textbf{k}}(\eta)$ obeys the equation of motion
\begin{equation}
    \partial^2_{\eta}\psi_k(\eta)+2aH\partial_\eta \psi_k(\eta)+\omega_k^2(\eta)\psi_k(\eta)=0,
\end{equation}
with an instantaneous frequency
\begin{equation}
    \omega_k^2=k^2+a^2(\eta)m^2+\frac{1}{6}(1-6\bar{\xi})a^2(\eta)R(\eta).
\end{equation}
\\
\\
To define an initial vacuum state at $\eta\rightarrow -\infty$ we impose the Bunch Davies conditions, which correspond to the standard Minkowski vacuum in the short wavelength limit:
\begin{equation}
    \psi_{k0}=\frac{1}{\sqrt{2k}},
\end{equation}
\begin{equation}
    \partial_\eta\psi_{k_0}(\eta)=-i\sqrt{\frac{k}{2}}e^{-ik\eta}.
\end{equation}
At early times, when the expansion is slow and modes are deep inside the horizon, these conditions ensure that the vacuum is uniquely defined.
\\
\\
The time dependence of the scale factor causes each mode to evolve non trivially, effectively mixing positive and negative frequency components. This mixing is elegantly captured by the Bogoliubov transformation, which provides a natural language for describing particle creation in curved spacetime.
\\
\\
As the universe evolves, the mode functions deviate from their initial form, and the vacuum defined at early times no longer coincides with that at late times. The relation between the two vacua can be expressed through the Bogoliubov coefficients $\alpha_k$ and $\beta_k$ which satisfy $|\alpha_k|^2-|\beta_k|^2=1$. The coefficient $|\beta_k|^2$ measures the number of particles produced in mode $k$.
\\
\\
The Bogoliubov coefficient $\beta_k$ quantifies how much a field mode is distorted by the expansion of the Universe. If the expansion is slow compared to the mode’s natural oscillation frequency, the mode evolves adiabatically and no particles are created. When expansion is rapid or highly oscillatory, positive and negative frequency components mix, giving a nonzero $\beta_k$. The quantity $|\beta_k|^2$ directly counts the number of particles created in mode $k$.
\\
\\
For each Fourier mode, the occupation number is obtained as
\begin{equation}
    |\beta_k|^2=\frac{1}{\omega_k}\left[\frac{1}{2}|\partial_\eta\psi_k|^2+\frac{1}{2}\omega_k^2|\psi_k|^2\right],
\end{equation}
where primes denote derivatives with respect to conformal time. This expression quantifies how strongly the mode deviates from a pure vacuum oscillation due to the changing background.
\\
\\
The total comoving particle number density is then,
\begin{equation}
    a^3n=\int a^3n_kdlog k,
\end{equation}
where $n_k$, defined by
\begin{equation} \label{a}
    a^3n_k=\lim_{n\rightarrow \infty}\frac{k^3}{2\pi^2}|\beta_k|^2
\end{equation}
The integrand $n_k$ represents the number of particles produced per logarithmic interval in momentum. In practice, the spectrum $n_k$ typically peaks around the scale where the violation of adiabaticity is strongest, usually near the end of inflation or during the first oscillations of the inflaton. Eqn.\eqref{a} describes particle creation in an idealized, non interacting setting. In realistic cosmological scenarios, however, the produced quanta can decay or scatter as the universe expands. To account for such processes, we include out of equilibrium corrections to the evolution of the Bogoliubov coefficients. In order to understand this processes, we plot the behavior of $n_k$ against $k$  in the limit $\eta\rightarrow -\infty$, for different values of case of $\xi$.
\begin{figure}[H]
    \centering
    
    \begin{subfigure}{0.48\textwidth}
        \includegraphics[width=\linewidth]{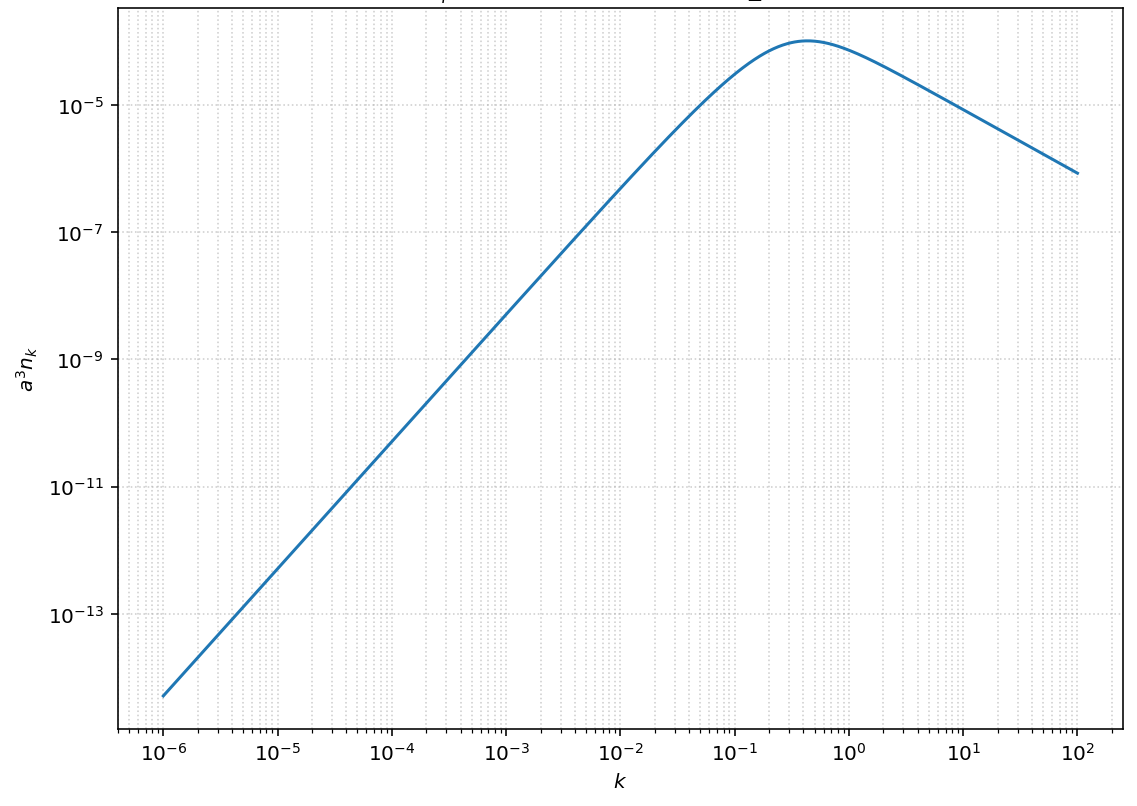}
        \caption{$\xi=0.02$}
        \label{Plot 4}
    \end{subfigure}
    \hfill
    \begin{subfigure}{0.48\textwidth}
        \includegraphics[width=\linewidth]{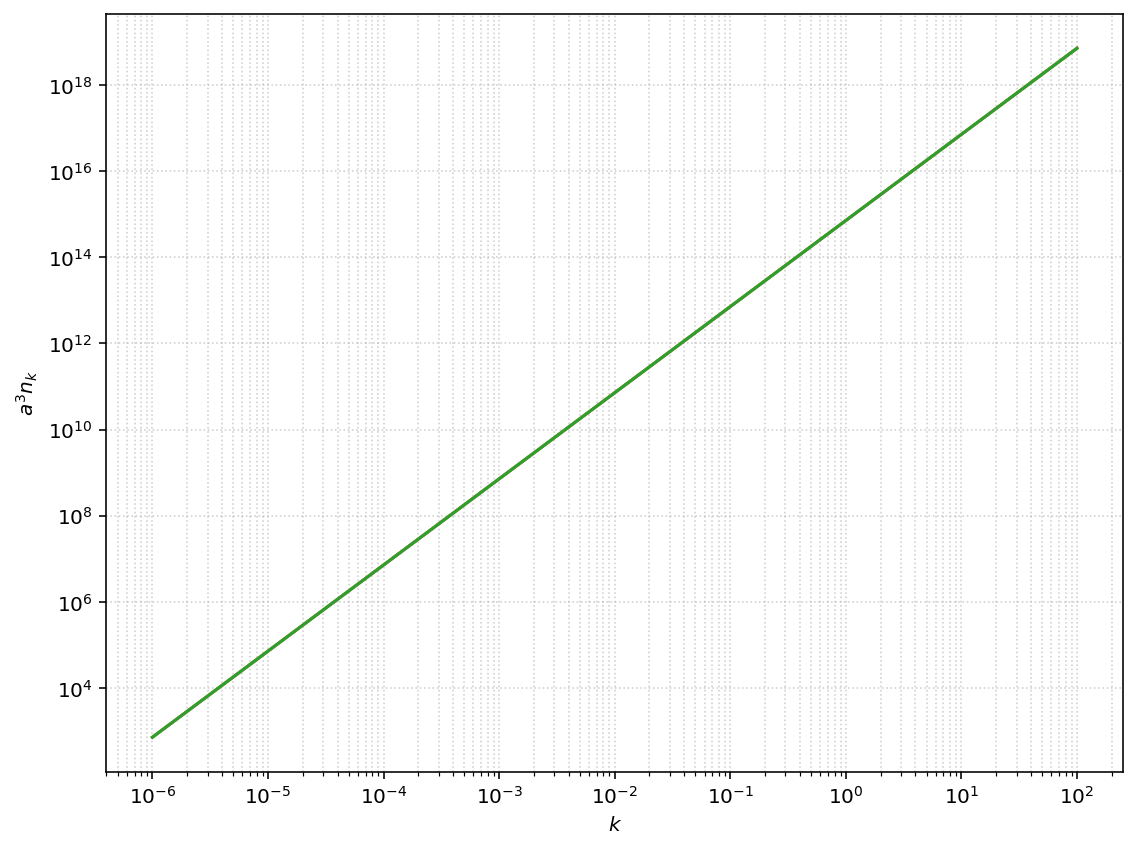}
        \caption{$\xi=1$}
        \label{Plot 5}
    \end{subfigure}

    \caption{$n_k$ against $k$  in the limit $\eta\rightarrow -\infty$, for different values of case of $\xi$.}
    \label{fig:3}
\end{figure}
In fig.\ref{Plot 4} we observe that for small $\xi$ ($\xi=0.02$), the curvature oscillations have modest amplitude, leading to particle production localized around the modes that most strongly violate the adiabaticity condition. This produces a characteristic peaked spectrum of $n_k$. While fig.\ref{Plot 5} shows that for large $\xi$ ($\xi=1$), curvature oscillations are much stronger, broadening the range of modes affected. The adiabaticity condition is violated for nearly all sub-horizon modes, producing a nearly linear rise of $n_k$. Thus, the shape of the spectrum reflects how efficiently curvature oscillations couple to different co-moving scales.
\\
\\
An intriguing consequence of geometric reheating is that the spectator scalar field $\psi$ can naturally play the role of a dark matter candidate \footnote{See for e.g works like \cite{Bertone_2005,Bertone_2018,Misiaszek_2024,vegetti2023stronggravitationallensingprobe,Battat:2024mcc,maxim2025quantum} for a concise review on dark matter}. Since $\psi$ is produced purely through gravitational particle production driven by the time dependent curvature, its abundance is largely insensitive to direct particle physics couplings with the visible sector. If $\psi$ stable or sufficiently long lived, the relic population generated during the reheating phase can survive to late times and contribute to the present dark matter density. The final abundance is then controlled primarily by the curvature coupling $\xi_{\psi}$, the expansion history during reheating, and the mass $m_\psi$. This mechanism realizes a purely gravitational production channel for dark matter, analogous to freeze in scenarios, but without requiring any non-gravitational interactions. As a result, $\psi$ provides a minimal and well motivated dark matter candidate, whose phenomenology is determined by the interplay between curvature induced particle production and cosmic expansion.
\\
\\
In particular, for $\xi >\frac{1}{6}$ (ensuring that Ricci scalar has a positive contribution to the effective mass) $\psi$ can naturally be interpreted as a late forming dark matter candidate (LFDM) candidate \cite{Sarkar_2015,Vattis_2019}. In LFDM scenarios, the dark matter abundance exists at early times but does not behave as cold, pressure less dark matter at recombination before transitioning to standard cold dark matter at later epochs \cite{Sarkar_2015}. Such a behavior reduces the effective dark matter density at recombination and modifies expansion history  prior to photon decoupling leading to a higher CMB inferred value of the Hubble constant, thereby alleviating the Hubble tension \cite{Vattis_2019}. Late forming dark matter can be realized through several mechanisms, including phase transition induced mass generation \cite{Shibuya_2022,ALANNE2014692}, dark radiation to matter conversion \cite{Ackerman_2009,Buen_Abad_2015} (which fails in alleviating Hubble tension, see for e.g \cite{mccarthy2023convertingdarkmatterdark})  or time dependent effective masses in the early Universe \cite{Amendola_2008,PhysRevD.108.083501}. 
\\
\\
In the present framework, the field $\psi$ provides a natural realization to late forming dark matter through its non minimal coupling to curvature. In the radiation dominated era, $R=0$, suppressing the curvature induced effective mass and rendering $\psi$ to effectively behave as radiation up until recombination. Moreover, during the radiation dominated epoch $\psi$,  with an negligible effective mass, scales as
\begin{equation}
 \psi\propto \eta^{-1}\propto a^{-1},
\end{equation}
yielding 
\begin{equation}
    \rho_{\psi}\propto (\partial_{\eta}\psi)^2 +\frac{1}{2}m^2_{eff}\psi^2\propto a^{-4}.
\end{equation} 
Furthermore,
\begin{equation}
    p_{\psi} \propto \frac{1}{3}(\partial_{\eta}\psi)^2 \propto \frac{p}{3},
\end{equation}
corresponding to that of radiation. As a result, the field does not contribute as cold dark matter at early times, reducing the effective dark matter density at photon decoupling. Following matter-radiation equality, the Ricci scalar becomes non zero, dynamically  generating an effective mass for $\psi$ through its curvature coupling (provided $\xi >\frac{1}{6}$). Once the effective mass exceeds the Hubble rate, the field begins to behave as  pressureless cold matter. In particular, the field scales as 
\begin{equation}
\psi \propto \eta^{-3} \propto a^{-3/2},
\end{equation}
which implies that,
\begin{equation}
    \rho_{\psi}\propto C\eta^{-6}+D\eta^{-8}\propto a^{-3},
\end{equation}
corresponding to that of matter. Here we have neglected the term which decays very fast and goes as $\eta^{-8}$.  Moreover,
\begin{equation}
    p\propto \eta^{-8}\propto 0,
\end{equation}
yielding a pressureless matter with $\omega=0$.
This delayed transition to cold dark matter behavior constitutes a minimal and geometrically motivated realization of late forming dark matter, which can alleviate the Hubble tension \cite{Di_Valentino_2021} by lowering the sound horizon. Furthermore, we see from Fig.\ref{Plot 5} that $\xi \sim O(1)$ curvature induced particle production during reheating does indeed generate a relic population whose late time energy density is consistent with the observed dark matter abundance, without requiring additional dark sector interactions.
\\
\\
Concretely, the sound horizon \cite{Hu_1996,Hu2002CMBReview} is given by
\begin{equation}
    r_s=\int_{z_{rec}}^{\infty} \frac{c_s}{H(z)}dz.
\end{equation}
Now, due to the additional scalar spectator particle which behaves like radiation until recombination $H(z)$ in our model in great than in the $\Lambda$CDM model. This inturn decreases the  sound horizon. Now, since the CMB \cite{Planck2020ResultsVI,2013ApJS..208...20B} measures 
\begin{equation}
    \theta=\frac{r_s}{D_A}
\end{equation}
very accurately, an lower $r_s$ implies a lower values of $D_A$ which implies a higher inferred value of $H_0$ from the CMB observation! A full numeric analysis as to what values of $\xi$ can fully alleviate Hubble tension \cite{Di_Valentino_2021} (possibly, even considering $S_8$ tension \cite{Di_Valentino2021}) will be carried forward alongside a detailed analysis on the structure growth of the LFDM in future works.
\\
\\
The above treatment assumes that particle production occurs in equilibrium, with modes freely evolving under the background expansion. In more realistic situations, the newly created quanta can interact or decay as the Universe evolves \cite{PhysRevLett.93.142002}. These effects can be incorporated phenomenologically by modifying the Bogoliubov evolution equation:
\begin{equation}
    \frac{d\beta_{k,outofeq.}}{dt}= \frac{d\beta_{k,eq.}}{dt}+\frac{\Gamma_{decay}}{2}\beta_{k}
\end{equation}
where $\Gamma_{decay}$ characterizes the rate of non-equilibrium interactions \cite{Berges_2004,boyanovsky1994nonequilibriumdynamicsphasetransitions}.
\\
\\
Consequently, the evolution of the comoving number density becomes
\begin{equation}
    \frac{d(n_ka^3)_{out of eq.}}{dt}=\frac{d(n_ka^3)_{eq}}{dt}+\Gamma n_ka^3.
\end{equation}
indicating that decays and scatterings can either enhance or suppress the total number of particles, depending on the sign and magnitude of $\Gamma$.
\\
\\
In physical terms, gravitational particle production acts as a quantum complement to the classical reheating process discussed earlier. Whereas the geometric reheating described in Sec.\ref{Section 2} converts inflaton energy into radiation through macroscopic curvature couplings, the mechanism here originates from microscopic quantum effects associated with a time dependent spacetime metric. Together, the two processes determine how efficiently the Universe can reheat when direct interactions between the inflaton and matter are absent. The combined effect sets the initial radiation temperature and, consequently, influences subsequent phenomena such as baryogenesis, dark matter production\footnote{Interested reader can see \cite{BeltranAlmeidaBernalRubioTenkanen2019} for the case that if cosmic inflation was driven by an electrically neutral scalar field stable on cosmological time scales then the field necessarily constitutes all or part of dark matter (DM) which is also an important line of investigation to take into account the non-perturbative nature of particle production during
reheating.}, and the generation of relic gravitational waves.

\section{Conclusions}\label{Conclusions}
In this work, we have explored the dynamics of reheating in inflationary models where the inflaton couples only gravitationally to other fields \cite{BassettLiberati1998,HaqueMaity2022}. Within this minimalist framework, referred to as geometric reheating, energy transfer from the inflaton to the Standard Model or other spectator sectors proceeds entirely through curvature effects, without requiring direct interaction terms \cite{BassettLiberati1998,HaqueMaity2022}. Such a mechanism is both economical and universal, as it relies solely on the gravitational sector, which must be present in any consistent theory of the early Universe.
\\
\\
Using a Higgs-like inflationary model \cite{chowdhury2025higgsinflationparticlefactory} with a non-minimal coupling $\xi\phi^2 R$, we derived the background equations of motion and examined how the curvature driven interaction can lead to the generation of a radiation bath after inflation. We found that the efficiency of reheating depends on the magnitude of $\xi$ and the post-inflationary equation of state parameter $\omega_{\phi}$. While the large values of $\chi$ can enhance the rate of energy transfer, renormalization group running of the Higgs quartic coupling $\lambda$ provides the flexibility to realize successful inflation and reheating even for moderate $\xi\sim 10$, thereby avoiding unitarity concerns associated with larger couplings.
\\
\\
In parallel, we analyzed gravitational particle production \cite{chowdhury2025higgsinflationparticlefactory,BirrellDavies1982,Fulling1989}, a purely quantum process by which fluctuations of spectator fields are excited due to the time dependent background curvature. Using the Bogoliubov formalism, we expressed the number density of produced quanta in terms of the mode functions and demonstrated how deviations from adiabatic evolution translate into particle creation. Furthermore, we encountered the interesting possibility where in the spectator particles can possible be interpreted as LFDM candidates \cite{Sarkar_2015,Vattis_2019}, effectively behaving as radiation in the radiation dominated phase but gradually transitioning into behaving as dark matter as the matter-radiation equality approaches. Moreover, extensions to include out of equilibrium effects show that decays or scatterings of these particles can further influence the total reheating efficiency \cite{Berges_2004,boyanovsky1994nonequilibriumdynamicsphasetransitions}. Also an interesting case arises when we use classical lattice simulations in 3+1 dimensions to study the interplay
between the resonant production of particles during preheating and the subsequent decay of
these into a set of secondary species \cite{RepondRubio2016}. This lattice-based extension to the non-linear regime when applied to this case can reveal some other important aspects which are worthwhile exploring in the future work.
\\
\\
Taken together, these results emphasize that gravitational and geometric processes alone can account for the transition from inflation to a radiation dominated Universe, even in the absence of explicit inflaton matter couplings. The combined effect of curvature induced reheating and gravitational particle production provides a natural and robust mechanism for initiating the thermal history of the cosmos.
\\
\\
Looking ahead, several extensions are worth pursuing. A detailed numerical analysis of the coupled inflaton radiation system would quantify the precise reheating temperature as a function of $\xi$ and $\Gamma_{\phi\phi\rightarrow RR}^{Rad}$. Incorporating fermionic or gauge spectator fields \cite{Greene2000FermionicPreheating} could reveal additional production channels relevant for baryogenesis and dark matter genesis. Finally, studying the associated spectrum of gravitational waves may offer observable imprints of geometric reheating, potentially connecting this early universe mechanism to future cosmological probes.
\section*{Acknowledgments}
The work of M. K. was performed in Southern Federal University with financial support of grant of Russian Science Foundation № 25-07-IF. The work of O.T. was supported in part by the Vanderbilt Discovery Doctoral Fellowship. We are grateful to Robert Brandenberger, Javier Rubio, Ayan Chakraborty, Jiahang Zhong and Ishan Swamy for their valuable discussions and comments.  Finally, we are also grateful to Prof. Pankaj Joshi, Meet Vyas, Aum Trivedi and peers at ICSC for their mentorship, advice, and continuous support.
\bibliographystyle{JHEP}
\bibliography{references}
\end{document}